# Soft Contract Verification for Higher-Order Stateful Programs

PHÚC C. NGUYỄN, University of Maryland, USA
THOMAS GILRAY, University of Maryland, USA
SAM TOBIN-HOCHSTADT, Indiana University, USA
DAVID VAN HORN, University of Maryland, USA

Software contracts allow programmers to state rich program properties using the full expressive power of an object language. However, since they are enforced at runtime, monitoring contracts imposes significant overhead and delays error discovery. *Soft contract verification* aims to guarantee all or most of these properties ahead of time, enabling valuable optimizations and yielding a more general assurance of correctness. Existing methods for static contract verification satisfy the needs of more restricted target languages, but fail to address the challenges unique to those conjoining untyped, dynamic programming, higher-order functions, modularity, and statefulness. Our approach tackles all these features at once, in the context of the full Racket system—a mature environment for stateful, higher-order, multi-paradigm programming with or without types. Evaluating our method using a set of both pure and stateful benchmarks, we are able to verify 99.94% of checks statically (all but 28 of 49, 861).

Stateful, higher-order functions pose significant challenges for static contract verification in particular. In the presence of these features, a modular analysis must permit code from the current module to escape permanently to an opaque context (unspecified code from outside the current module) that may be stateful and therefore store a reference to the escaped closure. Also, contracts themselves, being predicates written in unrestricted Racket, may exhibit stateful behavior; a sound approach must be robust to contracts which are arbitrarily expressive and interwoven with the code they monitor. In this paper, we present and evaluate our solution based on higher-order symbolic execution, explain the techniques we used to address such thorny issues, formalize a notion of behavioral approximation, and use it to provide a mechanized proof of soundness.

CCS Concepts: • **Software and its engineering** → **Formal software verification**; *Functional languages*;

Additional Key Words and Phrases: Higher-order contracts; symbolic execution

## 1 STATIC CONTRACT VERIFICATION IN A STATEFUL, HIGHER-ORDER SETTING

Software contracts [Findler and Felleisen 2002; Meyer 1991] allow programmers to provide rich specifications, using the full expressiveness of the host programming language, that are enforced dynamically. They have become a common mechanism for documenting and enforcing invariants

---

Authors' addresses: Phúc C. Nguyễn, University of Maryland, College Park, MD, USA; Thomas Gilray, University of Maryland, College Park, MD, USA; Sam Tobin-Hochstadt, Indiana University, Bloomington, IN, USA; David Van Horn, University of Maryland, College Park, MD, USA.

in many dynamically typed and higher-order languages [Austin et al. 2011; Disney 2013; Hickey et al. 2013; Plosch 1997; Strickland et al. 2012].

```
(define x 0)
(define/contract (f n)
  ((and/c int? (≥/c 0)) → (λ (n) (and/c int? (≥/c n))))
  (set! x (max x n))
  x)
```

To illustrate their use, consider the above function f written in Racket [Flatt and PLT 2010] that returns the largest natural number ever provided to it. Function f has a *dependent* contract enforcing that the function receives a single argument which must be a natural number and returns an integer no less than this argument.

While programmers may appreciate contracts for expressiveness and ease of use, as contracts are first-class values composable from arbitrary expressions, they have clear downsides: being checked dynamically delays error discovery and introduces non-trivial runtime overhead [Strickland et al. 2012]. Static verification of contracts eliminates both disadvantages, verifying program components, discovering errors up front, and turning previously expensive dynamic checks into strong static guarantees with no runtime overhead. It aids programmer confidence in software correctness while justifying the removal of runtime monitors which, in turn, can enable further optimizations that the presence of interposed branches and calls prevented. *Soft* contract verification aims to soundly overapproximate program behavior, verifying contracts where possible and gracefully degrading by allowing them to be enforced dynamically otherwise.

Verification of contracts in a stateful, higher-order, and dynamically-typed language presents unique challenges:

- The idioms of dynamic languages thwart simple verification methods such as type inference [Wright and Cartwright 1997], where programmers rely on dynamic type tests to justify uses of partial operations, with such tests being composed arbitrarily.
- Contracts are customarily used as enforcements at boundaries to ensure proper interaction between different components. Programmers tend not to write contracts for private functions, and rely on invariants established by interprocedural and path-sensitive reasoning. A compositional analysis is unlikely to succeed in verifying idiomatic dynamic programs without requiring heavy annotation from the programmer. Modularity, however, is crucial in scaling any analysis to a realistic code-base.
- Side effects complicate interactions between components through implicit communication channels. In particular, impure functions can escape the target module to be invoked an indeterminate number of times from an opaque context, possibly invalidating previously established invariants and triggering errors via interactions not possible in pure languages. Verification of effectful functions must soundly approximate such arbitrary interactions.
- Contracts themselves are arbitrary expressions capable of crashing, diverging, and modifying state, which prevents direct translation into pure functions and logical formulae for existing solvers.

Previous approaches to contract verification place greater restrictions on the language and contracts that limits applicability to a realistic programming environment. For example, they either rely on a static type system with contracts as function-level refinements [Knowles and Flanagan 2010; Vytiniotis et al. 2013; Xu 2012; Xu et al. 2009], or assume the language is pure [Nguyễn et al. 2014].

To advance this state of the art, we extend previous work on soft contract verification via higher-order symbolic execution, allowing mutable state in both concrete and symbolic functions. There were several crucial ingredients that made this possible. By employing a *path-condition* as standard in symbolic execution, our verification is inherently path-sensitive and capable of reasoning precisely on many idiomatic dynamically typed programs. By allowing symbolic values to be higher-order and effectful, we achieve modularity, enabling the omission of arbitrary program components to trade between precision and complexity of analysis, while internally incurring no precision loss from programming abstractions such as private functions and sub-modules. By tracking values that have escaped to unknown contexts and reasoning conservatively about locations that may be mutated at arbitrary points, we make the verification robust against unknown effectful functions. Finally, by relying on an extended operational semantics to encompass the behavior of higher-order and stateful program features, and employing an SMT solver for proving only first-order properties of restricted run-time structures, verification scales to a realistic programming language with provable soundness while remaining capable of taking advantage of SMT solvers for sophisticated theorem proving.

The primary technical delta from prior work on soft contract verification [Nguyễn et al. 2014] is the tracking of mutable functions and reasoning soundly about their possible effects. This required two extensions to our abstract semantics, we 1) maintained an overapproximation of the functions from transparent code that may have escaped to an unknown context, and then 2) extended our havoc semantics to simulate any sequence (of arbitrary order and length) of applying such functions at every point where control may escape to any unknown context.

**Contributions**

In this paper, we make the following four contributions:

(1) We give an extended operational semantics of a higher-order, stateful language that can be used for modular symbolic execution.
(2) We use a tunable abstraction process, enforcing termination of the analysis at some cost to precision, coupled with a formal notion of behavioral overapproximation in the presence of unknown higher-order, stateful values and a mechanically verified proof of soundness.
(3) We give a method for translating the symbolic execution history of a higher-order, stateful program into a pure, first-order formula, allowing the integration of a first-order SMT solver.
(4) We implement and evaluate a practical contract verifier for a significant subset of Racket.

## 2 EXAMPLES

This section explains the essential ingredients of our verification approach using examples. The approach is based on symbolic execution, which extends an existing language with *symbolic values*, each standing for an unknown, but fixed, concrete value. Symbolic execution explores a set of paths through a program, maintaining a *path condition* along each to remember facts which must be true about symbolic values on that specific path. Each path condition is a formula, characterizing a particular path, that is strengthened incrementally as execution passes through the conditional branches that separates this path from all others. By proving that some path conditions are contradictory and infeasible, the analysis is able to show that certain paths, and the errors along them, are unreachable.

If a symbolic execution were to explore all possible paths and terminate showing all errors to be unreachable, it would have performed a successful and complete verification that the program is statically free from run-time errors. In any real program, however, the number of distinct paths that may be explored is unbounded; in a traditional symbolic execution meant for finding bugs, imperfect coverage is just fine, for our purposes of static verification however, this unbounded set of paths must be soundly over-approximated. In addition, symbolic execution of higher-order functions requires simulating a program with unknown (opaque) functional values, i.e., we must reason about what happens when an unknown function is applied and the control path itself is unknown. In this section, we begin by discussing how symbolic execution can be used for program verification on a simple example, and then show its macro-expansion and a detailed view of our method on the program expanded to core forms (a smaller intermediate language). We then use further examples to discuss mutable state, effectful callbacks, and finitizing abstraction.

## 2.1 Path-sensitivity, SMT Solving, and Elaboration to Core Forms

Our first example demonstrates a function we can show is safe using path-sensitive reasoning over conditional control flow and first-order data—the basic building block of our verification. Being able to handle such programs is not unique to our process, but this example will allow us to illustrate the concepts underlying our approach before addressing higher-order values and mutable state. Function f on the left column of Figure 1 expects a pair of a real number and a string, and promises to return a real number. The function pattern-matches on its argument before applying partial operations such as str-len and division. A complete verification assures that not only is f correct with respect to its explicit contracts, but that the function's uses of partial operations (such as / and str-len) are also safe, and that pattern matching covers all cases.

Verification of f via symbolic execution begins by applying f to a fresh symbolic value for its argument. All paths through f start with a path condition already constrained so that f's argument matches the contract and is assumed to be a pair of a real number and string. Executing f's body would then non-deterministically follow each branch of the match form, because all of them could be possible. At each branch, symbolic execution of f proceeds with a strengthened path-condition, remembering the constraints on data that may be assumed down this particular branch. For example, all paths reaching the second match clause will record in their path conditions that x cannot be a pair where the first value is less than or equal to 1. In this way, path-conditions encode the invariants that ensure the absence of run-time errors, allowing us to prove that str-len is only applied on strings or that / is given a non-zero divisor. By calling out to a dedicated SMT solver, the analysis proves that, in the second clause, r must be both a real number and also not a real number less than or equal to 1, so therefore a non-zero real number. Down the implicit failure branch where Racket's match form would report an incomplete-pattern-match error, the path condition would report that x must be a pair of a real and string while also not a pair—a contradiction. The SMT solver will report that this path is infeasible and we have verified the match error can not occur. Ultimately, an exhaustive symbolic execution of just this component proves that f respects all its contracts (explicit or implicit), and no input can cause it to err.

Although the previous symbolic execution is straightforward for this example, it relies on knowledge of pattern-matching. Realistic programming languages can have a broad set of built-in features and, in the case of Racket, even user-definable syntax. Even match is simply a standard-library macro, with a pattern-matching facility that can be extended by user-defined macros. Building a verification process for a core language, after macro-expansion, allows better scaling of the process to large code bases using high-level language features, so long as the analysis can reason in terms of the fully-expanded intermediate language.

The program at the right column of Figure 1 shows the result of macro expanding function f into core Racket. The expanded program is more complicated code in terms of simpler forms and is idiomatic of dynamically typed languages in that correctness relies on path-sensitive properties such as aliasing and numerical bounds. For example, the match macro generates an alias $x_1$ for x, and then a thunk $k_2$ for exiting with an error if the match fails. Similarly, $r_1$, r, and $x_1$-car are aliases, and the program invokes thunk $k_1$, which performs a division on r only when its alias $r_1$ is greater than 1. Some thunks in the expansion of f would be unsafe if invoked from an arbitrary context; however, as this program uses them only in a correct way, our analysis proves that they are free from run-time errors. As our analysis constructs path conditions, refined across all dynamic checks, regardless of whether they are contracts specifically, no guard need be associated with these intermediate thunks. While in unexpanded Racket there are many language forms which may branch and refine the path condition, such as if, match, case, cond, for, do, etc, fully expanded Racket only has the if form. In addition, the many forms with complicated and unique control-flow behavior are compiled into administrative bindings, calls, returns, conditionals and continuations.

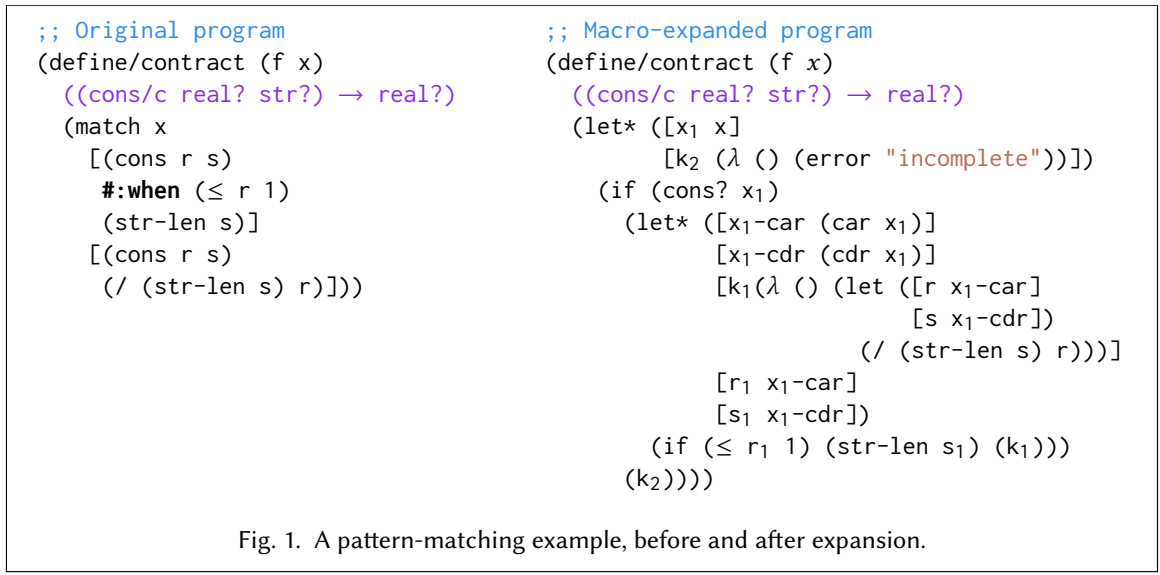

Fig. 1. A pattern-matching example, before and after expansion.

Verification of the expanded program proceeds in a manner similar to the original but needs to precisely track invariants about aliases and each closure's free variables. Variables in the analysis are bound to an *abstract* value that finitely approximates all possible runtime values (when it is entirely unknown, this is simply a • denoting a fully opaque, or unknown, value) paired with a *symbolic* value—a name that may be referenced in the path condition. The full mechanics of abstract and symbolic values is detailed and explained in Section 3. When execution reaches the let* form (a sequential let-binding form), it assigns to $x_1$ the value that x holds, which is an opaque value named x. Instead of being bound to a specific pair of concrete values, as it would be in any real execution, x is bound to a fully opaque abstract value (•) and a symbolic name, x (named for the original parameter), that is referenced in the path condition and constrained by it to be a pair of a real number and a string. At the let*-binding [$x_1$ x], this symbolic name propagates from x to $x_1$. The next binding evaluates the $\lambda$-term for a match error and assigns it to $k_2$. For closures, the analysis records both an abstract closure—the syntactic $\lambda$-term along with its abstract binding environment and the path condition at its creation (to constrain any free variables)—along with a symbolic name which is simply its syntactic $\lambda$-term.

The analysis then reaches a conditional to check if $x_1$ is a pair. In the "then" branch, the symbolic name x is assumed in the path condition to be a pair and any reference to variable $x_1$-car or $r_1$ returns a symbolic value named (car x), and for $x_1$-cdr or $s_1$, the symbolic name (cdr x). Symbolic names may refer to the value of any program expression—in this case, primitive operations. Most operations can be straightforwardly proven safe, except for the division in thunk $k_1$, where the path condition saved at the time of its creation did not assume that r was non-zero; this property was only recorded later before the application of $k_1$ when r's alias $r_1$ was assumed greater than 1. However, we treat the fact that $k_1$'s symbolic name is a $\lambda$-term as a signifier that the closure bound to k₁ was created within the caller. Assuming that the program has been $\alpha$-renamed, any free variable shared between k₁ and the current scope must therefore denote the same value. This justifies strengthening k₁'s saved path-condition with the caller's invariants, establishing that (car x) is greater than 1 and proving the division in k₁ safe. This is a precision-enhancing technique discussed further in Section 4.1. In all, this process proves that the expanded version of f is also free of run-time errors.

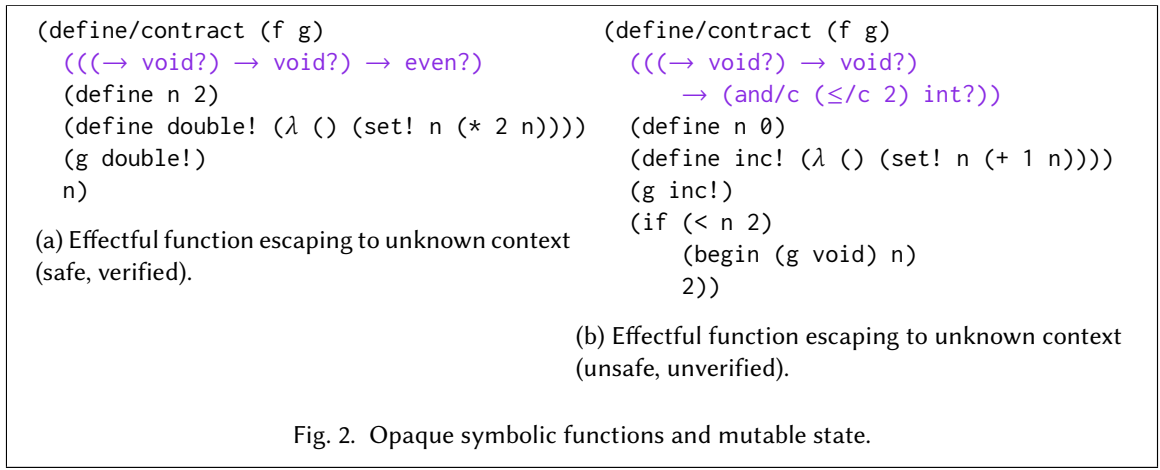

Fig. 2. Opaque symbolic functions and mutable state.

## 2.2 Effectful Callbacks and Mutated Location Tracking

The next two examples demonstrate the verification of a higher-order function with mutable state. Function f in Figure 2a takes as its argument any function g that accepts a callback, and promises to return an even integer. Internally, f defines a mutable variable n, defines a callback double! that modifies n to its double, and passes the callback to its argument g. Although g's behavior is arbitrary, it only causes modifications to f's local state n in a controlled way through double!. Our analysis is able to prove that n is always an even number, regardless of g's implementation, ensuring that f's result respects its contract.

To verify f in Figure 2a, we apply it to a fully opaque symbolic value. Within f's body, any reference to variable g returns a symbolic value named g. The semantics of higher-order contract monitoring wraps g in a contract promising that f only gives it a void-returning thunk, and ensuring that g only returns void [Findler and Felleisen 2002]. Upon applying g to double! and transferring control to g, the execution simulates arbitrary computation in g capable of affecting f's behavior and revealing its reachable errors. Specifically, g can both return a fully opaque symbolic value, and invoke (an arbitrary number of times) any function from f that has leaked to it (in this case, double!). Each update to n widens the value at n to approximate both old and new values. By choosing an appropriate domain of abstract values (e.g., one that preserves common predicates such as sign and parity tests) and providing a precise abstract interpretation for arithmetic over this domain, our analysis proves that n is always an even number regardless of how many times double! is invoked [Cousot and Cousot 1976, 1977]. Returning n can thus be shown to satisfy f's contract on its range and our symbolic execution verifies that f cannot be blamed for any violation of its contract. Section 4.1 describes the widening of values more generally.

Special care needs to be taken when a effectful callback escapes to an arbitrary context. For example, in Figure 2b, an opaque function g is invoked on inc!, a function that increments local variable n. The function, f, then tests to see if n remains strictly less than 2, in which case it invokes g on void and returns n, otherwise it returns 2. In the "then" branch, it may appear as though the call to (g void) may not effect n since the void function has no effect. However, an implementation of g that satisfies the contract can save the earlier reference to inc! and invoke it again many times when g is applied to void, invalidating any assumption about n's upper-bound.

### 2.3 Abstracting Symbolic Execution

Traditionally, symbolic execution is used to find potential errors, not to verify programs. Symbolic execution explores a number of program paths precisely but does not typically provide a terminating over-approximation of all possible program paths [Cadar et al. 2008; King 1976; Majumdar and Sen 2007; Sen 2007; Sen et al. 2005]. As described thus far, our process would not terminate on many programs. Consider factorial, shown in Figure 3: execution repeatedly unfolds factorial at each recursive call, applying the function to a fresh symbolic integer.

To ensure termination, we apply a well-studied method for systematically abstracting an operational semantics through finitization of dynamic program components [Van Horn and Might 2010]. The method yields many choices for finitizing the structure of an abstract machine [Might 2011] that are sound, and permits a method for tuning polyvariance to trade-off between precision and performance by changing the machine's allocation behavior [Gilray et al. 2016a]. Our instantiation of this framework approximates recurring values and path-conditions at different iterations of the same loop, summarizing repeated values and properties of data as loop invariants, while providing exact execution for loop-free program fragments. We describe our implementation in detail in Section 4.

```
(define/contract (factorial z)
  ((and/c int? (≥/c 0)) → (and/c int? (≥/c 1)))
  (if (≤ z 1)
      1
      (* z (factorial (- z 1)))))
```

Fig. 3. Factorial

In the case of factorial, execution branches on (≤ z 1), yielding two paths, and learns that factorial either returns 1 or multiplies z with the result of its recursive call (knowing that z is greater than 1). The recursive call similarly returns 1 in one of its branch, yielding (* z 1) as a result of the parent call to factorial. Through finitization of dynamically generated program components, the analysis reaches a fixed point, learning an over-approximation of factorial's behavior: it either returns 1, or the product of its argument z ≥ 1 with an integer no less than 1. Across all cases, the solver can verify that factorial satisfies its contracts.

## 3 STATIC VERIFICATION THROUGH SYMBOLIC EXECUTION

We present our symbolic execution using language $\lambda_S$, an untyped lambda calculus extended with mutable state and first-class higher-order contracts. The language's grammar is given in Figure 4.

Apart from standard values such as λ-abstractions (λ (x) e), natural numbers (n), and primitive operations (op), $\lambda_S$ includes symbolic values (•), each standing for an arbitrary value that is syntactically closed. (For example, • cannot stand for (λ (x) y) because y is free.)

Most forms are standard, such as variable reference (x), conditional (if e e e), and variable mutation (set! x e). We annotate each function application (e e)$^\ell$ with a source label $\ell$ to serve as a potential party to blame in case of a contract failure.

Contracts are first-class values in $\lambda_S$ and belong to the same syntactic category as expressions. Expression (e → (λ (x) e')) denotes a higher-order dependent contract, which is a pair of contract domain e and range "maker" (λ (x) e') that computes a range e' dependent on the argument

$$
\begin{aligned}
\text{[Expressions]} \quad & e \in \text{Exp} ::= u \mid x \mid (e\ e)^\ell \mid (\text{if}\ e\ e\ e) \mid (\text{set!}\ x\ e) \\
& \phantom{e \in \text{Exp} ::=} \mid (e \rightarrow (\lambda\ (x)\ e)) \mid (\text{mon}^\ell_{\ell'}\ e\ e) \\
\text{[Value Literals]} \quad & u \in \text{VExp} ::= (\lambda\ (x)\ e) \mid n \mid op \mid \bullet \\
\text{[Integers]} \quad & n \in \mathbb{Z} ::= \ldots \mid -2 \mid -1 \mid 0 \mid 1 \mid 2 \mid \ldots \\
\text{[Primitives]} \quad & op \in \text{Op} ::= \text{int?} \mid \text{proc?} \mid \text{zero?} \mid \text{flat-contract?} \mid \text{add1} \mid \ldots \\
\text{[Variables]} \quad & x, y \in \text{Var} = \langle \text{identifiers} \rangle \\
\text{[Labels]} \quad & \ell \in \text{Lab} = \langle \text{identifiers} \rangle
\end{aligned}
$$

Fig. 4. Syntax of $\lambda_S$.

bound to $x$ for each specific function application. For example,

$$(\text{int?} \rightarrow (\lambda\ (x)\ (\lambda\ (a)\ (\text{and}\ (\text{int?}\ a)\ (>\ a\ x)))))$$

is a contract for functions that map each integer to a greater integer.

Finally, the monitoring form $(\text{mon}^\ell_{\ell'}\ e\ e')$ denotes the dynamic enforcement of contract $e$ between expression $e'$ and its surrounding context, where $\ell$ is the party to blame if $e'$ produces a value that fails the contract, and $\ell'$ is to blame if the context consuming the result of $e'$ uses the result in a way that violates the contract. For example, $(\text{mon}^\ell_{\ell'}\ \text{int?}\ \text{add1})$ evaluates to a blame on $\ell$ for producing the function add1 instead of an integer, and the application $((\text{mon}^\ell_{\ell'}\ (\text{int?} \rightarrow \text{int?})\ \text{add1})\ \text{proc?})$ evaluates to a blame on $\ell'$ for supplying the primitive function proc? to a guarded function add1 expecting an integer.

```
(define (box x)
  (λ (msg)
    (match msg
      ['set-box! (λ (w) (set! x w))]
      ['unbox x])))
(define (set-box! b v)
      ((b 'set-box!) v))
(define (unbox b) (b 'unbox))
```

Fig. 5. Mutable box as closures.

Our language $\lambda_S$ is minimal, but models core features found in practical programming languages. For example, pattern matching can be expanded into simple conditionals as shown in Section 2.1, and mutable boxes can be modeled using closures and mutable variables, as demonstrated in Figure 5.

### 3.1 Semantics

We define the semantics of $\lambda_S$ using a reduction relation over machine states, $\varsigma$, each constituted of four components as shown in Figure 6.

At a high level, our abstract machine is just a closure-creating, store-passing interpreter, factored into several small-step rules and a single big-step rule that steps a configuration within an evaluation context (see Figure 7). We then instrument this machine with several extra components. In addition to a configuration and store, each state tracks a store-cache (mapping variables to locally precise values and their symbolic names) and a path condition (accumulating a conjunction of facts, in terms of these symbolic names, that characterizes the current execution path). We also include higher-order contracts as distinguished values, allow labeled blame objects to result from evaluation, and track strongly updated variables.

Soft Contract Verification for Higher-Order Stateful Programs    51:9

$$
\begin{array}{rl}
\text{[States]} & \varsigma \in \Sigma ::= (c, m, \phi, \sigma) \\
\text{[Closures]} & c \in \mathit{Conf} ::= a \mid (e, \rho) \mid (c\,c)^\ell \mid (\text{if } c\,c\,c) \mid (\text{set!}\,(x, \rho)\,c) \\
& \quad \mid (c \to ((\lambda\,(x)\,e), \rho)) \mid (\mathsf{mon}_\ell^\ell\,c\,c) \mid (\mathsf{rt}_x^{\vec{x}}\,s\,m\,\phi\,c) \\
\text{[Evaluation Contexts]} & \mathcal{E} \in \mathit{Ctx} ::= [\,] \mid (\mathcal{E}\,c)^\ell \mid (w\,\mathcal{E})^\ell \mid (\text{if } \mathcal{E}\,c\,c) \mid (\text{set!}\,(x, \rho)\,\mathcal{E}) \\
& \quad \mid (\mathcal{E} \to ((\lambda\,(x)\,e), \rho)) \mid (\mathsf{mon}_\ell^\ell\,\mathcal{E}\,c) \mid (\mathsf{mon}_\ell^\ell\,w\,\mathcal{E}) \\
& \quad \mid (\mathsf{rt}_x^{\vec{x}}\,s\,m\,\phi\,\mathcal{E}) \\
\text{[Answers]} & a \in \mathit{Ans} ::= w \mid \mathsf{blame}_\ell^\ell \\
\text{[Post-values]} & w \in \mathit{WVal} ::= (v, s) \\
\text{[Values]} & v \in \mathit{Val} ::= n \mid \mathit{op} \mid \bullet \mid \mathsf{Clo}(x, e, \rho, \phi) \mid \mathsf{Grd}(\alpha, \alpha) \mid \mathsf{Arr}_\ell^\ell(\alpha, \alpha) \\
\text{[Symbols]} & s \in \mathit{Sym} ::= e \mid \varnothing \\
\text{[Path-conditions]} & \phi \in \mathit{PC} = \mathcal{P}(\mathit{Exp}) \\
\text{[Address]} & \alpha \in \mathit{Addr} = \langle\text{any enumerable set, e.g. } \mathbb{N}\rangle \\
\text{[Store-cache]} & m \in \mathit{Cache} = \mathit{Var} \to (\mathit{WVal} + \{\varnothing\}) \\
\text{[Environments]} & \rho \in \mathit{Env} = \mathit{Var} \to \mathit{Addr} \\
\text{[Value stores]} & \sigma \in \mathit{Store} = \mathit{Addr} \to \mathcal{P}(\mathit{Val})
\end{array}
$$

Fig. 6. State components for symbolic execution.

*3.1.1 State Components.* Each machine state consists of four components:
(1) A closure ($c$) is either an answer ($a$), an expression ($e$) paired with an environment ($\rho$), or an inductively defined closure whose structure mimics that of plain expressions ($e$).
(2) A store-cache ($m$) is a finite map from each variable, $x$, in scope to either a value that $x$ must hold, or the special symbol $\varnothing$ indicating that $x$ may have been modified arbitrarily.
(3) A path condition ($\phi$) is a set (interpreted as a conjunction) of expressions assumed to have evaluated to true.
(4) A store ($\sigma$) that maps each address ($\alpha$) to a set of values ($\vec{v}$). For a standard concrete execution that is deterministic, each value set is a singleton. We generalize the store's range to be a set, however, to allow modeling non-determinism resulting from over-approximation of multiple execution paths.

Several further sub-components constitute these. Binding environments ($\rho$) map each variable in scope to an address ($\alpha$) in the value store. An answer ($a$) is either a post-value ($w$) or an error ($\mathsf{blame}_{\ell'}^\ell$) blaming component labeled $\ell$ for violating the contract with $\ell'$. A post-value ($w$) is a value ($v$) paired with a symbolic name ($s$) which relates this value to relevant constraints in the current path condition. The special symbol $\varnothing$ indicates the lack of a symbolic name to appear in the path condition.

A value ($v$) is either a number ($n$), primitive operator ($\mathit{op}$), function closure ($\mathsf{Clo}(x, e, \rho, \phi)$), higher-order contract ($\mathsf{Grd}(\alpha, \alpha)$), guarded function ($\mathsf{Arr}_\ell^\ell(\alpha, \alpha)$), or abstract value ($\bullet$). We use 0 to represent falsehood, and any other value to represent truth. For convenience, we assume zero? is a total predicate in $\lambda_S$ that tests for falsehood, and nonzero? is its complement. Similarly, flat-contract? tests if a value is usable as a flat contract (e.g. either a primitive function or a lambda), and dep-contract? tests if a value is a dependent contract.

The grammar for evaluation contexts ($\mathcal{E}$) follows that of closures ($c$) and enforces a standard left-to-right call-by-value semantics. In particular, we evaluate functions before arguments, and contracts before monitored expressions.

The semantics is mostly standard with the addition of a store-cache and a path-condition that tracks path-sensitive information about locations and symbolic values, respectively. For example, an entry x ↦ (•, (+ y n)) in the store-cache means that location x holds the symbolic value (+ y n), and expression (> x n) in the path-condition indicates the constraint on symbolic values x and n. The name x, when it appears as a symbol, refers to the value first bound to location x when x is in scope.

*3.1.2 Reduction relation.* We define the small-step operational semantics using reduction relation ($\longmapsto$), which defines the evaluation of a machine state as a sequence of atomic steps. The relation ($\longmapsto$) is defined as the context-closure of relation ($\longmapsto_v$) over redexes, as shown in figure 7. We present each aspect of this reduction relation in turn.

Some rules (e.g. *[Grd], [AppClo], [MonFun]*) involve allocating a value in the store at some address $\alpha$—we leave the choice of address allocation open, because any allocation results in a sound over-approximation of the standard concrete semantics [Gilray et al. 2016a; Van Horn and Might 2010]. As we will see in Section 3.5, different allocation choices decide whether the semantics is a traditional bug-finding symbolic execution or static verification with different trade-offs between precision and termination.

*Distribution of environment into sub-expressions.* Rule *[Distr]* in Figure 8 shows the reduction of closures of the form $(e, \rho)$, where $e$ contains one or more sub-expressions. The (partial) meta-function distr shows the straightforward definition of distributing environments into sub-expressions.

*Values.* Rule *[Lit]* in Figure 9 shows the reduction for value literals. Each of these expressions evaluates to an answer determined by meta-function lit. The definition of lit is straightforward for base values. For each $\lambda$-abstraction, the meta-function saves the current path-condition to remember invariants about free variables, in addition to the environment as standard. Rule *[Grd]* shows the reduction of a higher-order contract once its domain is evaluated: the reduction steps to a contract object with both its domain and range-maker components allocated in the store.

*Variable referencing and mutation.* Figure 9 also shows reduction rules for referencing and mutating variables. Rule *[Var]* references the value at variable $x$ by performing a lookup in the store-cache ($m$) as well as in the store ($\sigma$). As shown in the definition of relation lookup, we either use the cache result if the cache indicates a definite hit, or look up in the store as standard if the cache indicates that the variable has stopped being tracked precisely. Rule *[Set]* strongly updates the store-cache and weakly updates the store.

*Conditionals.* The last rules in Figure 9 shows reduction for conditionals. If the evaluated condition is plausibly non-0, execution steps to the "then" branch of the conditional and refines the path-condition to reflect the new assumption (rule *[CondTrue]*). Here, relation feasible($\phi$, $op$, $w$, $\phi'$) guards the rule, ensuring that value $w$ could possibly satisfy predicate $op$, given the current invariants in path-condition $\phi$, and remembering the branch condition as an assumption of a strengthened path-condition $\phi'$. If the condition is plausibly 0, execution steps to the "else" branch of the conditional and refines the path condition to reflect the inverse assumption (rule *[CondFalse]*). When both branches are plausible, execution non-deterministically steps to both, each with the appropriately strengthened path condition.

$$\frac{(c_1, m_1, \phi_1, \sigma_1) \longmapsto_v (c_2, m_2, \phi_2, \sigma_2)}{(\mathcal{E}[c_1], m_1, \phi_1, \sigma_1) \longmapsto (\mathcal{E}[c_2], m_2, \phi_2, \sigma_2)}$$

Fig. 7. Reduction on states.

$$((e, \rho), m, \phi, \sigma) \longmapsto_v (c, m, \phi', \sigma) \quad [Distr]$$
$$\text{if} \quad distr(e, \rho) = c$$

where  $distr((e_1\ e_2)^\ell, \rho) = ((e_1, \rho)\ (e_2, \rho))^\ell$
$distr((\text{if } e\ e_1\ e_2), \rho) = (\text{if } (e, \rho)\ (e_1, \rho)\ (e_2, \rho))$
$distr((\text{set! } x\ e), \rho) = (\text{set! } (x, e)\ (e, \rho))$
$distr((e \rightarrow (\lambda\ (x)\ e')), \rho) = ((e, \rho) \rightarrow ((\lambda\ (x)\ e'), \rho))$
$distr((\text{mon}^\ell_{\ell'}\ e\ e'), \rho) = (\text{mon}^\ell_{\ell'}\ (e, \rho)\ (e', \rho))$

Fig. 8. Distribution of environment.

$$((u, \rho), m, \phi, \sigma) \longmapsto_v (\text{lit}(u, \rho, \phi), m, \phi, \sigma) \quad [Lit]$$
$$(((v, s) \rightarrow ((\lambda\ (x)\ e), \rho)), m, \phi, \sigma) \longmapsto_v ((\text{Grd}(\alpha_1, \alpha_2), \varnothing), m, \phi, \sigma') \quad [Grd]$$
where  for some addresses $\alpha_1, \alpha_2 \quad \sigma' = \sigma \sqcup [\alpha_1 \mapsto v, \alpha_2 \mapsto \text{Clo}(x, e, \rho, \phi)]$
$$((x, \rho), m, \phi, \sigma) \longmapsto_v (w, m, \phi, \sigma) \quad [Var]$$
where  $\text{lookup}(\sigma, \rho, m, x) \ni w$
$$((\text{set! } (x, \rho)\ (v, s)), m, \phi, \sigma) \longmapsto_v ((1, \varnothing), m', \phi, \sigma') \quad [Set]$$
where  $m' = m[x \mapsto (v, s)] \quad \sigma' = \sigma \sqcup [\rho(x) \mapsto v]$
$$((\text{if } w\ c_1\ c_2), m, \phi, \sigma) \longmapsto_v (c_1, m, \phi', \sigma) \quad [CondTrue]$$
where  $\text{feasible}(\phi, \text{nonzero?}, w, \phi')$
$$((\text{if } w\ c_1\ c_2), m, \phi, \sigma) \longmapsto_v (c_2, m, \phi', \sigma) \quad [CondFalse]$$
where  $\text{feasible}(\phi, \text{zero?}, w, \phi')$

where  $\text{lit}(n, \_, \_) = (n, n) \quad \text{lit}(op, \_, \_) = (op, op)$
$\text{lit}(\bullet, \_, \_) = (\bullet, \varnothing) \quad \text{lit}((\lambda\ (x)\ e), \rho, \phi) = (\text{Clo}(x, e, \rho, \phi), (\lambda\ (x)\ e))$

where  $\text{lookup}(\sigma, \rho, m, x) \ni w$, if $m(x) = w$
$\text{lookup}(\sigma, \rho, m, x) \ni (v, \varnothing)$, if $m(x) = \varnothing$ and $v \in \sigma(\rho(x))$

Fig. 9. Reduction on values, variables, mutation, and conditionals.

*Contract monitoring.* Figure 10 shows reduction rules for monitoring contracts.

Rule *[MonFlat]* shows the straightforward monitoring of a flat contract—the contract is simply applied on the value as a predicate. If the value passes this predicate, it is returned as-is; otherwise, a blame on the party providing the value is raised.

$$((\text{mon}_{\ell'}^{\ell}, w' \ w), m, \phi, \sigma) \longmapsto_v ((\text{if } (w' \ w) \ w \ \text{blame}_{\ell'}^{\ell}), m, \phi', \sigma) \quad \text{[MonFlat]}$$
where $\text{feasible}(\phi, \text{flat-contract?}, w', \phi')$

$$((\text{mon}_{\ell'}^{\ell}, w' \ w), m, \phi, \sigma) \longmapsto_v ((\text{if } (\text{proc? } w) \ \text{Arr}_{\ell'}^{\ell}(\alpha_1, \alpha_2) \ \text{blame}_{\ell'}^{\ell}), m, \phi', \sigma') \quad \text{[MonFun]}$$
where $\text{feasible}(\phi, \text{dep-contract?}, w', \phi')$ $(v, \_) = w$ $(v', \_) = w'$
and for some addresses $\alpha_1, \alpha_2$, $\sigma' = \sigma \sqcup [\alpha_1 \mapsto v', \alpha_2 \mapsto v]$

Fig. 10. Reduction on contract monitors.

Rule *[MonFun]* shows the monitoring of a higher-order contract, which first performs a first-order check ensuring the target is indeed a function, blaming the party providing the value if it is not. If the value is indeed a function, monitoring saves the higher-order contracts, the function being checked, along with the blame parties into a guarded function to perform checks at each subsequent application, following the semantics of monitoring higher-order contracts. The details of applying a guarded function are described later in application rule *[AppArr]*.

*Application.* Figure 11 shows reduction rules for application. Values that can be used as functions in $\lambda_S$ are primitive operations, closures, and guarded functions, whose applications are shown in rules *[AppPrim]*, *[AppClo]*, and *[AppArr]*, respectively. Applying an opaque value results in two possibilities covered in rule *[AppOpq]*.

Application of primitive operations rely on relation $\delta$ as in rule *[AppPrim]*. We generalize $\delta$ to a relation instead of meta-function to express non-determinism in the presence of symbolic values. The result's symbolic name is created through meta-function ap, which reconstructs the application, except returning $\varnothing$ if either arguments is $\varnothing$.

Rule *[AppClo]* governs application of a closure and allocates the argument in the store at an address $\alpha$. Because some variables have new visible bindings, the store-cache and path-condition need to be updated. The target closure's parameter ($x$) now referes to a distinct location from the caller's $x$ (if it exists), so it receives a fresh entry in the store-cache pointing to the argument's value as well as $x$ as the symbolic name. The closure's free variables, on the other hand, may or may not refer to the same locations as they do at the call site, so their store-cache entries are simply invalidated and treated conservatively. The closure application reduces to the function body with the extended environment and store as standard, but also saves the caller's store-cache, path-condition, and symbolic name to reinstate upon a return.

Rule *[AppArr]* describes the application of a guarded function, which is decomposed into the monitoring of the argument against the contract's domain with reversed blame parties, followed by the application of the function under guard, whose result is in turn monitored against the computed contract range. If the function's guarding contract is not a concrete higher-order contract, it decomposes into opaque domain and range contracts.

Rule *[AppOpq]* describes two non-deterministic cases that result from applying an opaque function. Because a blame on an unknown program component is an irrelevant analysis result, we ignore unknown functions that introduce errors of their own. A well-behaved function, on the other hand, interacts with the rest of the program in limited ways: it either returns a value, or if its argument $v$ is a function, applies $v$ to some value. Because each application of $v$ may modify program state, and the effect of each application (*e.g.*, whether it triggers an error) may depend on the program state resulting from applying a stateful function (either $v$ itself or another function that has previously escaped from transparent code to unknown code), arbitrary repetition of this application needs to be considered. We therefore maintain a set of values that have escaped to

$$(((op, s')\ (v, s))^\ell, m, \phi, \sigma) \longmapsto_v ((v', \mathsf{ap}(s', s)), m, \phi, \sigma) \quad [AppPrim]$$
where $\quad v' \in \delta(\sigma, op, v)$

$$(((\mathsf{Clo}(x, e, \rho, \phi), s_f)\ (v, s))^\ell, m, \phi, \sigma) \longmapsto_v ((\mathsf{rt}_x^{\overrightarrow{y}}\ s'\ m\ \phi\ (e, \rho')), m', \phi', \sigma') \quad [AppClo]$$
where $\quad \overrightarrow{y} = dom(\rho) \quad w = (v, s) \quad m' = m[x \mapsto (v, x)]\overrightarrow{[y \mapsto \varnothing]} \quad s' = \mathsf{ap}(s_f, s)$
and for some address $\alpha, \quad \rho' = \rho[x \mapsto \alpha] \quad \sigma' = \sigma \sqcup [\alpha \mapsto v]$

$$((\mathsf{Arr}^\ell_{\ell'}(\alpha_c, \alpha_f)\ w)^\ell, m, \phi, \sigma) \longmapsto_v \quad [AppArr]$$
$$((\mathsf{mon}^\ell_{\ell'}\ w_r\ (w_f\ (\mathsf{mon}^{\ell'}_\ell\ w_d\ w))), m, \phi, \sigma)$$
where $\quad \mathsf{Grd}(\alpha_d, \alpha_r) \in \sigma(\alpha_c) \quad v_d \in \sigma(\alpha_d) \quad v_r \in \sigma(\alpha_r) \quad v_f \in \sigma(\alpha_f)$
$w_d = (v_d, \varnothing) \quad w_r = (v_r, \varnothing) \quad v_f = (v_f, \varnothing)$

$$(((\bullet, s')\ (v, s))^\ell, m, \phi, \sigma) \longmapsto_v \begin{cases} (w_\bullet, m, \phi', \sigma') \\ ((w_\bullet\ (w_i\ w_\bullet)), m, \phi', \sigma') \end{cases} \quad [AppOpq]$$
where $\quad \mathsf{feasible}(\phi, \mathsf{proc?}, (\bullet, s'), \phi') \quad \sigma' = \sigma \sqcup [\alpha_\bullet \mapsto \{v\}]$
$v_i \in \sigma'(\alpha_\bullet) \quad w_\bullet = (\bullet, \varnothing) \quad w_i = (v_i, \varnothing)$

$$((w'\ w)^\ell, m, \phi, \sigma) \longmapsto_v (\mathsf{blame}^\ell_\Lambda, m, \phi', \sigma) \quad [AppErr]$$
where $\quad \mathsf{feasible}(\phi, \mathsf{nonproc?}, w', \phi')$

Fig. 11. Reduction on application sites.

$$((\mathsf{rt}_x^{\overrightarrow{y}}\ s'\ m'\ \phi'\ (v, s)), m, \phi, \sigma) \longmapsto_v ((v, s''), m'', \phi, \sigma) \quad [Ret]$$
where $\quad m'' = m[x \mapsto m'(x)]\overrightarrow{[y \mapsto \varnothing]} \quad s'' = \varnothing \text{ if } s = \varnothing, s' \text{ otherwise}$

Fig. 12. Reduction on function returns.

unknown code at a special address $\alpha_\bullet$, and upon invocation of an opaque function, we emulate an arbitrary number of invocations of escaped code before finally returning an opaque value as a result. The first case of rule *[AppOpq]* approximates all possible returns from the unknown function by returning an opaque value. The second case of rule *[AppOpq]* approximates all possible applications of the unknown function by applying any value from the transparent code that has "leaked" into the unknown code (including the argument from the latest opaque application, $v$), and then passes the result back into an opaque function.

To illustrate how rule *[AppOpq]* uncovers errors, consider the example in Figure 13 where execution discovers a potential division-by-0 error in function f when passed to an unknown context in an execution branch where f is applied thrice. In another example, given in Figure 14, the first unknown context cannot discover any error in the function app which flows to it, because there is no possible error. However, its result, stored at variable h, can potentially reference any value that has escaped to an unknown context. After the the effectful

```
(let ([f (let ([x -3])
           (λ (_)
             (set! x (+ 1 x))
             (/ 1 x)))])
  (• f))
```

Fig. 13. Havoc stateful callback.

function inc! flows to the unknown context, the variable n has potentially been modified to 0. Application of h then soundly discovers the potential error in app by invoking app again.

```
(let* ([n -3]
       [inc! (λ (_) (set! n (add1 n)))]
       [app (λ (_) (/ 1 n))]
       [h (• app)])
  (• inc!)
  (h 0))
```

Fig. 14. Escaped stateful callback.

In Section 3.4, we provide a precise definition of behavioral over-approximation and show that these rules are sufficient to soundly approximate the application of an unknown function with arbitrary code. Although a naïve implementation of rule *[AppOpq]* is impractical, we employ several optimizations to enable verification of realistic programs presented in Section 4.

Lastly, in *[AppErr]*, applying a potentially non-functional value blames the party performing the application.

*Restoring context.* Figure 12 shows rule *[Ret]* for returning a value to the caller. The store-cache entry for the distinct bound-variable $x$ is restored, while entries for free variables are simply invalidated. In addition, the value receives the symbolic name from the caller's scope.

### 3.2 Primitive operations

Figure 15 shows a definition of select primitive operations extended to symbolic values.

Although many primitives such as add1 are partial, we define *op* in $\lambda_S$ to be the *unsafe* versions of the primitives, which are total functions that always successfully return a value. We therefore assume that references to primitives are appropriately guarded with contracts (e.g. add1 would be guarded with (int? → (λ (_) int?))) and that programmers have no direct access to unsafe primitives.

The definition preserves precision for concrete arguments and returns an opaque value otherwise.

### 3.3 Path-condition satisfiability

Effective verification relies on precise proving of infeasible path-conditions to eliminate implausible blames and avoid exploration of spurious paths. While simple properties such as implication and exclusion between type-like predicates are easy to check, more sophisticated properties such as arithmetics take more work to implement efficiently. Making good use of an existing SMT solver can reduce implementation effort without giving up the ability to prove rich invariants. Unfortunately, most SMT solvers only support first-order formulae, which are a significant gap from higher-order effectful expressions.

$$\begin{array}{lll}
\delta(\sigma, \text{int?}, v) & \ni 0 & \text{if } v \neq n \\
\delta(\sigma, \text{int?}, v) & \ni 1 & \text{if, } v = n \text{ or } v = \bullet \\
\delta(\sigma, \text{add1}, n) & \ni n + 1 & \\
\delta(\sigma, \text{add1}, v) & \ni \bullet & \text{where, } v \neq n
\end{array}$$

Fig. 15. Primitive operations.

We overcome this issue by making the following observation: runtime monitoring, even of higher-order values, only requires checking a first-order property at any given point in the program execution. Therefore, contract verification ultimately reduces to proving implications between first-order properties. We rely on the operational semantics to account for the execution of a program, while accumulating first-order invariants in the path-condition to be able to prove necessary properties. Call outs to the solver can be seen a precision optimization that prunes infeasible paths of execution. We present a method to translate the path-condition into first-order formulae such that unsatisfiable formulae implies an infeasible execution path. The formulae's satisfiability

$$\{\!|\cdot|\!\} : \phi \to \textit{Formula} \qquad\qquad \{\!|\cdot|\!\} : e \to \textit{Term}$$

$$\{\!|e\ldots|\!\} = \wedge\, (\texttt{istrue}\, \{\!|e|\!\})\ldots \qquad\qquad \{\!|n|\!\} = \texttt{Int}\, n$$

$$\{\!|op|\!\} = \texttt{Op}\, \text{unique}(op)$$

$$\{\!|\cdot,\cdot|\!\} : op \times e \to \textit{Term} \qquad \{\!|(\lambda\, (x)\, e)|\!\} = \texttt{Lam}\, x,\ \text{where}\ x\ \text{is fresh}$$

$$\{\!|\texttt{add1}, e|\!\} = \texttt{Int}\, (1 + \texttt{unbox\_int}\, \{\!|e|\!\}) \qquad \{\!|(op\, e)|\!\} = \{\!|op, e|\!\}$$

$$\ldots \qquad\qquad \{\!|e|\!\} = x,\ \text{where}\ x\ \text{is fresh}$$

$$\{\!|op, e|\!\} = x,\ \text{where}\ x\ \text{is fresh}$$

Fig. 17. Translation of path-conditions and expressions into first-order formulae.

can be solved by an existing SMT solver such as Z3 [Moura and Bjørner 2008] or CVC4 [Barrett et al. 2011].

The target formulae uses a unitype embedding V of the source language's dynamic type system as shown in Figure 16. Source language values are encoded in the solver by pairing together a run-time type tag and an integer denoting the identity of the value: for source integers, it is just the integer itself; for all other kinds of values such operators, functions, etc., it just distinguishes values.

```
type V = Int (unbox_int: ℤ)
       | Op  (op_id: ℤ)
       | Lam (lam_id: ℤ)
       | ...
function istrue v = (v ≠ Int 0)
```

Fig. 16. Datatype encoding for SMT solver.

Figure 17 shows the translation. The main translation takes a path-condition $\phi$ and produces a formula stating properties about run-time values. It straightforwardly asserts that the translation of each term in the path-condition is not `Int 0`. Unsatisfiability of this formula would imply an infeasible execution path.

The translation of each expression $e$ produces a term of sort $V$ in the logic. Base values are straightforwardly mapped to those in the logic, while the translation of functions such as primitives and lambdas merely retain the type tag. The translation uses a fresh id for each lambda literal, essentially existentializing the translated value. Primitive operations that have a correspondence in the logic are translated as is. For operations and expressions that do not have obvious translations, we simply existentialize the result, as seen in the default cases of $\{\!|\cdot|\!\}$ and $\{\!|\cdot,\cdot|\!\}$.

### 3.4 Soundness

This section proves the symbolic execution semantics is sound—that it discovers any possible blame. Specifically, given a program with holes, if *any* instantiation of the holes causes a blame on a label from the incomplete program, then running the program (with holes) under the symbolic semantics discovers the same blame (Theorem 3.2).

To prove our soundness theorem, we first define what it means for an incomplete program to *approximate* a complete one, then through a preservation lemma, we show reduction preserves this approximation.

Figure 18 shows important rules for the approximation relation between expressions and state components. We describe only the important rules and defer to the appendix for full definitions. The base case of the derivation involves the instantiation of holes (•), where each hole either stands for a literal base value, or a syntactically closed $\lambda$-abstraction whose body ($e^\bullet$) does not contain further holes and only contains a label ($\ell_\bullet$) distinct from any label from the transparent part of the code. Approximation rules for expressions arise from the straightforward structural induction.

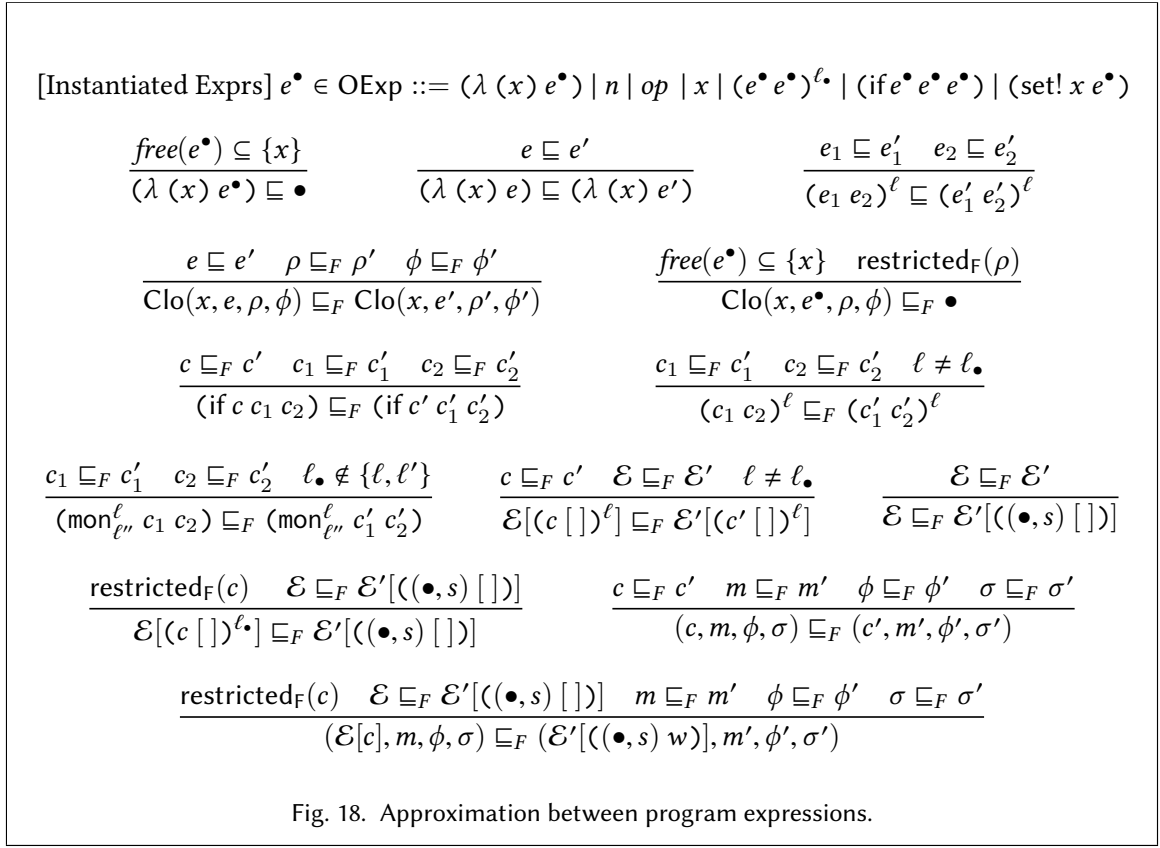

Fig. 18. Approximation between program expressions.

The approximation between state components is indexed by an *abstraction map F* from each address in the instantiated component to one in the approximating component. The approximation between closures is only established between closures whose corresponding sub-components are approximating, or between an opaque value and a closure whose control component is purely instantiated, and an environment component that only maps to "unknown" addresses simulated by the special address $\alpha_\bullet$ (enforced by map $F$). Predicate restricted$_F(\cdot)$ restricts state components to only contain addresses that are simulated by $\alpha_\bullet$. Approximation between evaluation contexts also include structural and non-structural cases: inserting transparent "frames" into the holes of both evaluation contexts preserves approximation, and an inner opaque application context approximates any number of insertions of purely instantiated "frames". Finally, approximation between states ($\varsigma$) is either established structurally through component-wise approximation, or non-structurally where an opaque application approximates a state whose evaluation context and top frames are purely instantiated.

We also define the multi-step standard reduction ($\longmapsto\!\!\!\!\rightarrow$) as the reflexive-transitive closure of the standard reduction ($\longmapsto$).

LEMMA 3.1 (REDUCTION PRESERVES APPROXIMATION). *If $\varsigma_1 \sqsubseteq_F \varsigma_1'$ and $\varsigma_1 \longmapsto \varsigma_2$, then there exists $\varsigma_2'$ and $F'$ such that $\varsigma_2 \sqsubseteq_{F'} \varsigma_2'$ and $\varsigma_1' \longmapsto\!\!\!\!\rightarrow \varsigma_2'$.*

PROOF. By case analysis on the reduction $\varsigma_1 \longmapsto \varsigma_2$ and approximation $\varsigma_1 \sqsubseteq_{F'} \varsigma_1'$. We defer to the appendix for the full proof, and link to our mechanization, written in Lean. Most cases are straightforward and $\varsigma_2'$ approximates $\varsigma_1'$ in lock step. The main complication comes from applying symbolic function, where the instantiated state $\varsigma_1$ transfers control to purely instantiated code ($e^\bullet$), and the symbolic state steps with [*AppOpq*]. When this occurs, the same state that succeeds [*AppOpq*] continues to approximate an arbitrary number of states that $\varsigma_1$ steps to, as long as $\varsigma_1$'s control comes from instantiated code. By approximation, instantiated code can only transfer to transparent code through returning, or applying one of the "leaked" values approximated by those at $\alpha_\bullet$, which [*AppOpq*] soundly simulates. □

With the established small-step soundness of $\lambda_S$, we prove that running an incomplete program $e'$ approximates the result of running any of its full instantiation $e$. We define a helper meta-function $\mathsf{load}(e) = ((e, \{\}), \{\}, \emptyset, \{\})$ that loads the initial state of a program.

THEOREM 3.2 (BLAME SOUNDNESS). *If $e_1 \sqsubseteq e_1'$ and $\mathsf{load}(e_1) \longmapsto\!\!\!\!\twoheadrightarrow (\mathcal{E}[\mathsf{blame}_{\ell'}^\ell], m, \phi, \sigma)$, where $\ell \neq \ell_\bullet$, then there exists $\mathcal{E}'$, $m'$, $\phi'$, and $\sigma'$, such that $\mathsf{load}(e_1') \longmapsto\!\!\!\!\twoheadrightarrow (\mathcal{E}'[\mathsf{blame}_{\ell'}^\ell], m', \phi', \sigma')$.*

PROOF. The proof proceeds by rule-induction on the derivation of ($\longmapsto\!\!\!\!\twoheadrightarrow$) in the concrete error trace. The base case (reflexive) vacuously holds. The inductive case (transitive) holds by lemma 3.1, where for each single reduction step ($\longmapsto$) on the concrete state, the abstract state continues to approximate the concrete state in zero or more steps. □

Corollary 3.3 follows from theorem 3.2, stating a practical implication for verification: if an incomplete program is safe under the symbolic execution, no instantiation of the program can causes a blame on any part of it.

COROLLARY 3.3 (VERIFIED COMPONENTS CANNOT BE BLAMED). *If $\mathsf{load}(e) \not\longmapsto\!\!\!\!\twoheadrightarrow (\mathcal{E}[\mathsf{blame}_{\ell'}^\ell], m, \phi, \sigma)$ for any label $\ell$ appearing in $e$, then there is no instantiation $e' \sqsubseteq e$ such that $\mathsf{load}(e') \longmapsto\!\!\!\!\twoheadrightarrow (\mathcal{E}'[\mathsf{blame}_{\ell'}^\ell], m', \phi', \sigma')$.*

PROOF. The corollary holds as the contrapositive of Theorem 3.2. □

## 3.5 From symbolic execution to verification

Traditionally, symbolic execution was used for finding bugs in programs, and not for static verification, because, as originally formulated, it does not provide a terminating overapproximation that guarantees the absence of run-time errors. To turn bug-finding symbolic execution into a verification process, we employ an existing method for turning an operational abstract-machine semantics into an overapproximation through a systematic finitizing of machine components [Van Horn and Might 2010]. Thus far, we have allowed the closure to grow without bound, and have left value addresses ($\alpha$) allocation unspecified. A fresh allocation at each transition will yield a concrete execution (assuming no opaque values), but a different allocation strategy that repeatedly reuses addresses, and joins (conflates) values at those addresses within the store, results in an approximation of multiple execution traces. Indeed, any allocation policy is sound [Might and Manolios 2009]. This is because all possible abstract allocators are consistent simulations of the concrete allocator because the latter always allocates a fresh address. This leaves the choice of allocation as a central "tuning knob" for adjusting the analysis's precision using any desired degree of polyvariance or context sensitivity [Gilray et al. 2016a]. We also transform the evaluation contexts into explicit *continuations*, and store-allocate continuations at function boundaries and permit two continuations to become conflated at a single *continuation address*; this redefines each continuation to be a sequence of intraprocedural frames paired with a continuation address for the current invocation. Although the continuation allocator may also be adjusted arbitrarily, recent work has shown that in order to achieve precise call-and-return matching at no asymptotic cost to analysis complexity, the choice of continuation address should be fixed as $(e, \rho)$ where $e$ and $\rho$ are the call target's control and environment respectively [Gilray et al. 2016b].

The property we desire for path-sensitive contract verification is that the analysis should only approximate values at different iterations of the same loop, and provide exact execution otherwise. We therefore instrument execution with another component recording the set of control transfers from each source's location to a target's function body. Each such set is an abstraction for a family of traces that differ from one another only by the number of iterations through the same loops. By pairing each syntactic component (e.g. variables) with this set in the allocated address, we obtain an abstraction that meaningfully summarizes program components such as values, continuations, and path conditions with probable cycles resulting from loops. Although this allocation strategy does not guarantee a worst-case polynomial time analysis, (*i.e.*, a loop-free exponential time program would result in exponential time analysis), it tends to give good precision and analysis time for real programs. In addition, modularity helps mitigate the potential worst cases as the user can always break a large program into smaller modules to verify separately.

Finally, we perform a standard global-store widening that weakens the correlation between the store for values, continuations, and path conditions, and other machine components by lifting these to the top level collecting semantics and maintaining the store as the least-upper-bound of all stores visited across all paths. If some of the precision lost during this transformation is needed, it may be regained through the use of a more precise allocation strategy. Strategies, such as Shivers' time-stamp algorithm [Shivers 1991], may also be used to avoid revisiting a machine configuration until the global store is updated.

## 4 IMPLEMENTATION AND EVALUATION

This section discusses practical improvements in implementing contract verification, and examines our verifier's analysis time and precision on a variety of benchmarks.

### 4.1 Practical improvements

*Richer abstract values and widening of base values.* The formalism in Section 3.1 uses only a single abstract value (•) that represents "any value". In our implementation, we enrich each such abstract value to carry a *refinement set* containing predicates it is known to have satisfied. For example, $\bullet^{\{\text{int?,positive?}\}}$ denotes an abstract positive integer. These refinements provide our analysis semantics a way to short-circuit a call to the SMT solver. For our experiments, we restrict the refinement set to predicates on base types, along with those that syntactically appear in the program (e.g., user-defined contracts), as this provides a balance between precision and convergence. When two different values share the same address, they are widened to an abstract value whose refinement set they both satisfy. In addition, primitives such as addition and multiplication are extended to operate precisely on such abstract values.

*Avoid re-running escaped values.* Rule *[AppOpq]* in Section 3.1 over-approximates all behavior triggered by the unknown part of the program, but is expensive when implemented naïvely due to an ever-growing set of leaked values to be applied at each opaque application. To reduce this cost, we memoize the result of applying each leaked value by the portion of the value store that can potentially affect its behavior, and only re-run a leaked value if the memoized portion of the store has lost precision since that value was last run. This is especially effective at speeding-up mostly-functional programs since pure functions do not depend on or modify mutable state, and are thus only explored once.

*Inter-procedural path-sensitivity.* Path-sensitivity across function boundaries is crucial for verifying programs with predicates that are abstracted arbitrarily. We achieve this improvement by augmenting the semantics with a *memo-table* that explicitly records information about each application's results and the corresponding path-conditions at each result. At any point in the execution, the memo-table maintains an over-approximation of properties that must hold for each application's arguments and results. Each entry in the memo-table is then translated into an uninterpreted first-order function along with formulae about arguments and results for observed cases. These additional formulae yield more constraints that allow eliminating more spurious paths.

*Sharing invariants for provably same locations.* Application rule [AppClo] conservatively uses the callee's saved path-condition, and returning rule [Ret] conservatively invalidates store-cache entries for the callee's free-variables. In the restricted case when the target function is known to share the same free-variables as its caller, properties pertaining to variables shared between a call site and the invoked closure can be combined, strengthening instead of invalidating the saved path condition. Our semantics maintains the invariant that a value's symbolic name is a $\lambda$-term only when it is instantiated in the same scope as its caller (by inspection of the reduction rules, only rule *[Lit]* produces a $\lambda$-term as a symbolic name). Identical variable names in these cases imply identical dynamic locations (assuming that the program has been $\alpha$-renamed). We therefore achieve additional precision in these particular cases by sharing the path-condition's constraints and the store-cache entries for those locations between callers and callees that are provably the same.

*Let-aliasing.* Finally, realistic programming languages allow storing intermediate results in variables, and some programs may rely on reasoning through aliases such as those introduced by macro-expansion as shown in Figure 1. In a language with `let`-aliasing, we simply allow the store-cache to initialize each `let`-bound variable to the first value that flowed to them, effectively canonicalizing the symbolic names for values at `let`-bound variables. For example, in the following safe function f (Figure 19), looking up both y and z within the function body would give a value with symbolic name (car x). Any test on y would give information about z and vice versa. As previously noted in Section 3.1, the name x appearing in symbolic names means the value first bound to location x and not the location x itself. Any subsequent modification to location x only modifies the store-cache and does not invalidate the path-condition.

### 4.2 Implementation

We extend the core semantics described in Section 3.1 to a practical implementation that verifies contracts in full Racket programs. By handling core forms directly, and invoking the macro expander to desugar all others, the tool is able to work on significant Racket programs. Compared to the formalism, the implementation provides significant extensions.

```
(define (f x)
  (let* ([y (car x)]
         [z y])
    (when (integer? y)
      (set! x #f)
      (set! y #f)
      (add1 z))))
```

Fig. 19. Stateful program with let-aliasing.

First, base values are much richer, including the full numeric tower and values such as strings and symbols. Second, we support datatypes such as pairs, mutable boxes, mutable vectors, and user-defined structs with mutable fields. Third, we support additional contract combinators including disjunction, conjunction, recursion, etc, and monitor contracts using *indy* instead of *lax* semantics as presented in rule *[AppArr]* in Section 3.1 for complete contract monitoring with correct blame parties [Dimoulas et al. 2011]. Finally, we support multiple return values and arbitrary function arities, resulting in several additional possible errors.

The implementation is available at https://github.com/philnguyen/soft-contract.

### 4.3 Evaluation

To evaluate the tool's effectiveness, we collect benchmarks from several lines of previous work including soft typing for Scheme [Wright and Cartwright 1997], occurrence type-checking [Tobin-Hochstadt and Felleisen 2010], higher-order model checking [Kobayashi et al. 2011], and symbolic execution [Nguyễn et al. 2014; Tobin-Hochstadt and Van Horn 2012]. In addition, we verify other realistic libraries collected from different sources. We include summarized results for small benchmarks from previous work on verification of pure programs to show the lack of regress. Different benchmark suites each emphasize different aspects of verification. The occurrence-typing suite includes small programs whose correctness heavily relies on reasoning about path-sensitivity and aliasing, which is common in untyped programs. The hors suite includes many higher-order recursive programs, where safety relies on inter-procedural reasoning. The benchmarks snake, tetris, and zombie are moderately-sized programs with expressive contracts that were collected from an introductory programming course. Finally, remaining benchmarks are existing Racket libraries and programs collected from multiple sources, written in the full Racket language with imperative features. In total, our benchmarks are comprised of 86.7% stateful benchmarks (by lines of code) and 13.2% pure functional benchmarks.

Table 20 show benchmark results. Line counts do not include empty and commented lines. The number of checks is a static count of the number of safety checks, including those in primitives and user-specified contracts, which could be eliminated if proven correct. Although the number of checks seems unintuitively high, it reflects the reality of safe dynamic languages. For example, each call site checks if it is applying a function, and each arithmetic operation checks that it is passed numbers. We also include the number of checks resulting from user-written contracts in parentheses on the right of columns Checks, False Pos, and True Pos. True and false positives are determined through manual inspection. Finally, verification time is measured in seconds.

Our results show that our tool not only can verify almost all contracts and works for many interesting programming patterns, with reasonable analysis time even for large programs. For example, slatex initializes mutable boxes with sentinel values (e.g. #f), then updates them in a type-consistent way afterwards (e.g. proper non-empty list). Our analysis proves all these uses safe. Another program, nucleic2-modular, uses vectors to emulate records with fields having different types, and passes data to many higher-order and partially applied functions, and our analysis verifies that all the indices are in-bounds and updates and references are type-consistent. In addition, the analysis's modularity makes it practical, where the programmer can break a large program into multiple modules to verify separately. For example nucleic2 is originally a closed program taken from a standard Scheme benchmark suite, and has literal vectors contributing to more than half of the code. Although a good stress test, closed programs are not the focus of our modular verification. Therefore, we abstracted out the input data and verified the computation, demonstrating the intended use case. Finally, among the potential errors reported, some are genuine bugs, as in slatex, such as applying an operation expecting a pair where an empty list is possible. We also fix these errors and report the result as slatex*.

Imperative benchmarks include some realistic programs we cannot fully verify. Further inspection reveals that false positives come from a few specific programming patterns.

First, the tool cannot yet reason about invariants established by a module that controls the instantiation of certain structures and maintains strong invariants about all instances (for example, that a "node" is always part of a non-empty proper tree). This is seen in the ring-buffer and leftist-tree programs. Because our semantics is conservative in assuming that opaque values can come from anywhere, we cannot precisely reason about this pattern. A simple and efficient solution permitting reasoning about this idiom is an important goal for future work. Modified versions (`ring-buffer*` and `leftist-tree*`) with deep structural contracts enable the analysis to succeed in verifying the programs.

| Program | Lines | Checks | Time (s) | False Pos | True Pos |
|---|---:|---:|---:|---:|---:|
| soft-typing | 108 | 656 (37) | 2.064 | 0 (0) | 0 (0) |
| hors | 266 | 2,194 (119) | 4.828 | 2 (2) | 0 (0) |
| occurence-typing | 87 | 647 (51) | 2.423 | 0 (0) | 0 (0) |
| snake | 142 | 1,232 (96) | 2.485 | 0 (0) | 0 (0) |
| tetris | 259 | 2,390 (200) | 6.578 | 0 (0) | 0 (0) |
| zombie | 235 | 1,049 (39) | 3.000 | 0 (0) | 0 (0) |
| fector | 110 | 388 (20) | 4.578 | 4 (0) | 0 (0) |
| hash-srfi-69 | 290 | 1,920 (97) | 4.125 | 1 (1) | 0 (0) |
| leftist-tree | 102 | 916 (25) | 0.656 | 8 (0) | 0 (0) |
| leftist-tree* | 110 | 918 (25) | 0.562 | 0 (0) | 0 (0) |
| morsecode | 185 | 1,013 (12) | 4.968 | 0 (0) | 0 (0) |
| nucleic2-modular | 884 | 6,621 (11) | 88.453 | 1 (0) | 0 (0) |
| nucleic2-modular* | 889 | 6,644 (11) | 84.062 | 0 (0) | 0 (0) |
| ring-buffer | 51 | 353 (19) | 0.563 | 8 (0) | 0 (0) |
| ring-buffer* | 58 | 354 (19) | 0.438 | 0 (0) | 0 (0) |
| slatex | 2,300 | 11,633 (2) | 1,213.650 | 2 (0) | 6 (0) |
| slatex* | 2,305 | 11,693 (2) | 1,217.850 | 2 (0) | 0 (0) |
| TOTAL | 8.381 | 49,861 (785) | 2,641.283 | 28 (3) | 6 (0) |

Fig. 20. Benchmark Results

Second, our analysis does not yet precisely verify invariants established by effectful functions, as seen in the `fector` benchmark. In this module, several functions rely on an operation `reroot!`, presented in Figure 21 to guarantee that the content of the mutable box is a vector. Because most data-structures in verification arise from unknown sources, and these data structures after passing through recursive contracts are cyclic (to approximate all possible values inhabiting the contracts), addresses typically point to one or more abstract values abstract multiple concrete values; preventing symbolic execution from performing strong updates to such addresses. More precise abstraction and reasoning for mutable recursive data-structures is a second goal for future work.

```
(define (reroot! fv)
  (match (unbox fv)
    [(list i x fv*)
     (reroot! fv*)
     (let ((v (unbox fv*)))
       (let ((x* (vector-ref v i)))
         (vector-set! v i x)
         (set-box! fv v)
         (set-box! fv* (list i x* fv))))]
    [_ (void)]))
(reroot! fv)
; use of `vector-length` not verified
(vector-length (unbox fv))
```

Fig. 21. reroot! example from fector

## 5 RELATED WORK

Our work builds on existing approaches to static contract verification via symbolic execution. We relate our current contributions to these efforts and then more broadly to work on verification in higher-order settings.

### 5.1 Symbolic execution

Symbolic execution simulates a program's evaluation on symbolic values which are unknown and may stand in for a number of possible concrete values. Path conditions—formulas unique to a particular sequence of branches—constrain these symbolic variables and denote infeasible runs where contradictory. In first-order settings, symbolic execution has a mature and well-investigated methodology [Cadar et al. 2008, 2006]; in higher-order settings however, it remains an active area of ongoing research. In a higher-order setting, where a concrete value may be a first-class function, a variety of sound choices exist for modeling the application of an opaque (symbolic) function which do not exist in first-order languages [Tobin-Hochstadt and Van Horn 2012].

There is also a general difference in motivation; while most applications of symbolic execution involve bug finding and code auditing, our focus is on its use for modular program verification and static contract checking.

### 5.2 Static contract verification

We have built on prior work [Nguyễn and Van Horn 2015; Nguyễn et al. 2014; Tobin-Hochstadt and Van Horn 2012] that develops static contract verification as a (higher-order) symbolic execution of untyped functional programs (in this case, Racket). Previous work following this approach only handles pure functions, and while robust for untyped functional programs, it falls down in the presence of even well-encapsulated mutable state and other non-functional idioms. Further, the implementation presented in that work handled only a small subset of Racket.

Another approach [Xu 2012; Xu et al. 2009] embeds dynamic monitors into the target program and simplifies them away using compiler techniques and a specialized symbolic engine. This approach of symbolic simplification may be applicable to untyped programs; however, a crucial pass used in this approach, dubbed logicization, requires type annotations in order to translate program expressions into a first-order logic (FOL). A similar method for Haskell [Vytiniotis et al. 2013] leverages a denotational semantics that can be mapped onto first-order logic; this is both dependent on type information and on the pure call-by-name semantics of Haskell.

Contract verification in the setting of first-order contracts is also more restricted, and its investigation more mature. A prominent example is the work on verifying C# contracts done in the Code Contracts project [Fähndrich and Logozzo 2011] and the Spec# system [Barnett et al. 2011; Müller and Ruskiewicz 2011], with which, contract counter examples can be generated and explored using a debugger.

Our approach allows higher-order dependent contracts and mutable state, does not assume types to guide the verification process, supports blame, and verifies runtime type safety in addition to richer contracts as part of the same process. In addition, the aforementioned type-based approaches assume explicit monitoring of recursive calls which allow the use of contracts as inductive hypotheses in such calls. Our approach permits this as well, but remains flexible enough to accommodate Racket's semantics which does not monitor recursive call sites.

### 5.3 Refinement type checking

Refinement type systems permit the inclusion of logical propositions within type annotations and represent another approach to statically stating and proving richer properties of programs—as such, there is meaningful overlap with contract verification. Refinement type systems either restrict the expressivity of type refinements so that checking is decidable [Freeman and Pfenning 1991], or they permit arbitrary refinements, as do contracts in Racket, and use a general-purpose solver in the attempt to discharge refinements [Knowles and Flanagan 2010; Rondon et al. 2008; Vazou et al. 2013]. When a refinement cannot be discharged, a system may reject the program as a whole [Rondon et al. 2008; Vazou et al. 2013], or, as in the case of hybrid type checking [Knowles and Flanagan 2010], it may residualize a runtime check to dynamically enforce each unverified refinement. Manifest contracts [Greenberg et al. 2010; Gronski and Flanagan 2007] equip static types with contracts as refinements, verifing contracts either statically via subtyping, or using a dynamic cast. Manifest contracts have also been extended to algebraic data and mutable state [Sekiyama and Igarashi 2017; Sekiyama et al. 2015], including stateful contracts. Residualized runtime checks correspond to our approach of *soft* contract verification which degrades gracefully, removing only those contracts which are verified. Unlike our approach, manifest contracts and hybrid type checking require type annotations and only permit predicates on base types; while our approach extends to dependent contracts, no mechanism currently exists for mixing flat and higher-order specifications in refinement types. Furthermore, contract evaluation may become stuck, diverge, or have side effects, while refinements are more restricted. Special care must be taken where refinements themselves contain a potentially failing cast [Greenberg et al. 2010; Knowles and Flanagan 2010]. Dependent JavaScript [Chugh et al. 2012a,b] supports expressive refinements for stateful JavaScript programs, including sophisticated dependent specifications. Unfortunately, this approach relies on extensive type annotations and whole-program analysis.

### 5.4 Higher-order model checking

Higher-order model checking is also applicable to verification problems in this setting. This approach proceeds by compiling a target program into a higher-order recursion scheme (HORS)—these are essentially programs in the simply-typed $\lambda$-calculus, with finitely inhabited types, that generate unbounded trees representing all possible program evaluation paths. While HORS generalizes finite-state and pushdown systems, its model checking problem remains decidable while in this ideal setting of simply-type $\lambda$-calculus and finite base types [Kobayashi 2009b; Kobayashi et al. 2010; Ong 2006]; however, there remains a significant gulf between this and real-world language features. Other work has broadened the applicability of this approach to cases [Neatherway et al. 2012], to untyped languages [Kobayashi and Igarashi 2013; Tsukada and Kobayashi 2010], and to infinite data domains such as integers and algebraic datatypes [Kobayashi et al. 2011; Ong and Ramsay 2011]. The complexity of higher-order model checking is $n$-EXPTIME-hard [Kobayashi and Ong 2009] but practical progress has lead to engines which can handle checking some "small but tricky ... functional programs in under a second" [Kobayashi 2009a].

While our approach tackles untyped, higher-order, stateful programs with sophisticated real-world language features, higher-order model checking is restricted to small, pure code snippets using a more restricted set of features. In addition, our approach allows programmers to add dynamically enforced program invariants via contracts and dispatch them gradually while the HORS approach only supports assertions on first-order data which must all be verified. Our approach also permits verification in the presence of unknown library functions (not only base values), a crucial allowance for modular program verification. Our evaluation demonstrates that our tool can verify many of the "small but tricky" examples checked in the HORS literature.

### 5.5 Broadly related static analysis

Separation logic [Birkedal et al. 2008; O'Hearn et al. 2001; Reynolds 2002] provides a framework for reasoning compositionally about the abstract effect of computations—we do not. More broadly, summarization-based and bottom-up approaches aim to produce modular and compositional analyses of program components. Our work takes a more operational view and does not summarize the effects of known functions (instead it simulates them under an approximation). It does aim, however, to summarize the effects of arbitrary unknown functions. This is done by operationally tracking which heap locations an unknown function has access to. This process does bear some resemblance to a frame rule or localization technique: if an unknown function does not have access to a location, it cannot have an effect on it. As our evaluation shows, this is sufficient to verify the prepoderance of contracts in our benchmark suite.

As the target language is functional, benchmarks use mutable data sparingly and usually in a well-encapsulated manner. In addition, contracts in Racket usually enforce properties of a component's input and output data, not about its side effects. For these reasons as well, we were able to obtain good reasults without separation logic. We are not automatically verifying full functional correctness properties of programs that make heavy use of pointer manipulation. While it is possible to express, for example, a linked-list reversal algorithm that does pointer swapping, our approach is unlikely to do a good job proving it correct. Rather, we target type- and dependent-type-like properties expressed as contracts on higher-order, imperative programs.

The trace semantics described in [Laird 2007] is also related to our system. Both use non-determinism and opaque values to model the behavior of components in a context that includes mutation. This work uses denotational semantics and addresses fundamentally distinct concerns. This system yields a sound and complete representation of a program's operational behavior, whereas our system yields a computable approximation aimed at static analysis applications.

Other notions of abstraction for addresses can also be used with our approach; for example, a recency abstraction [Balakrishnan and Reps 2006] tracks abstraction cardinality (singleness) [Might and Shivers 2006] in addition to a distinguished non-approximate heap location. If a heap location is known to be single, our approach may be able to yield improved results where a strong-update is enabled ahead of a contract.

## 6 CONCLUSION

Contracts allow programmers to enforce sophisticated invariants within their code using the power and expressiveness of the host language. However, this flexibility comes at a cost to runtime efficiency and without any compile-time assurance of correctness. Soft contract verification offers a remedy—by attempting to verify contracts statically; where a contract can be verified, its code may be removed, permitting optimization of the underlying program, and the program property it had enforced at runtime will have been proven for all possible executions. We demonstrate that symbolic execution may be extended to support higher-order languages with mutable state in modular fashion, permitting arbitrary interaction with unknown external components. This extension to opaque (unknown and potentially stateful) functions requires us to make subtle but crucial choices; for example, accounting for the possibility of a known function escaping permanently to an opaque context. Our approach scales to the full Racket programming language and our evaluation shows that our tool can verify more than 99.9% of dynamic checks across a suite of realistic (14% pure and 86% stateful) benchmarks.

### ACKNOWLEDGMENTS

We are grateful to the anonymous reviewers of OOPSLA 2017 and POPL 2018 and to the anonymous artifact evaluation reviewers of POPL 2018; their constructive feedback has improved this research. This work is supported in part by NSF grant 1618756, the NSA Science of Security lablet program, and the UMD Basili Postdoctoral Fellowship. We thank William Kunkel for help developing our soft contract verification tool and Ben Greenman for being an early adopter and frequent bug reporter.

## A SOUNDNESS

This appendix establishes the formal soundness of the symbolic execution semantics, justifying its use for program verification. The mechanized proof of soundness is available in the supplemental material accompanying this paper.

### A.1 Approximation

The approximation relation ($\sqsubseteq$) between components is indexed by an *abstraction map* ($F$) from each address in the instantiated component to one in the approximating component.

Structural cases are straightforward. In non-structural cases where the right-hand side is •, some components in the left-hand side are enforced by predicate restricted$_F(\cdot)$ that all controls ($e^\bullet$) are purely instantiated, and all addresses only reference either values instantiated by unknown code or those from the set of "leaked" values from the known code. The latter property is established by making $F$ map all "unknown" addresses to the special address $\alpha_\bullet$, which holds • and the set of all "leaked" values from the transparent code.

### A.2 Proof

Lemma A.1 (Reduction preserves approximation).
*If $\varsigma_1 \sqsubseteq_F \varsigma_1'$ and $\varsigma_1 \longmapsto \varsigma_2$ then there is some $\varsigma_2'$ and $F'$ such that $\varsigma_2 \sqsubseteq_{F'} \varsigma_2'$ and $\varsigma_1' \longmapsto\!\!\!\!\!\rightarrow \varsigma_2'$*

Proof. By case analysis on the derivation of $\varsigma_1 \sqsubseteq_F \varsigma_1'$ and $\varsigma_1 \longmapsto \varsigma_2$.

- Case $\mathcal{E}[(e, \rho)] \sqsubseteq_F \mathcal{E}'[(e', \rho')]$ where the distr relation holds: Next states continue to approximate by rule *[Distr]*.
- Case $\mathcal{E}[(u, \rho)] \sqsubseteq_F \mathcal{E}'[(u', \rho')]$: Next states continue to approximate by rule *[Lit]*

$$\frac{e \sqsubseteq_F e'}{(\lambda\ (x)\ e) \sqsubseteq_F (\lambda\ (x)\ e')} \qquad \overline{n \sqsubseteq_F n} \qquad \overline{op \sqsubseteq_F op} \qquad \overline{x \sqsubseteq_F x}$$

$$\frac{e_1 \sqsubseteq_F e_1' \quad e_2 \sqsubseteq_F e_2' \quad \ell \neq \ell_\bullet}{(e_1\ e_2)^\ell \sqsubseteq_F (e_1'\ e_2')^\ell} \qquad \frac{e_1 \sqsubseteq_F e_1' \quad e_2 \sqsubseteq_F e_2' \quad e_3 \sqsubseteq_F e_3'}{(\text{if}\ e_1\ e_2\ e_3) \sqsubseteq_F (\text{if}\ e_1'\ e_2'\ e_3')}$$

$$\frac{e \sqsubseteq_F e'}{(\text{set!}\ x\ e') \sqsubseteq_F (\text{set!}\ x\ e')} \qquad \frac{e_1 \sqsubseteq_F e_1' \quad e_2 \sqsubseteq_F e_2'}{(e_1 \to (\lambda\ (x)\ e_2)) \sqsubseteq_F (e_1' \to (\lambda\ (x)\ e_2'))}$$

$$\frac{e_1 \sqsubseteq_F e_1' \quad e_2 \sqsubseteq_F e_2' \quad \ell_\bullet \notin \{\ell', \ell''\}}{(\text{mon}^\ell_{\ell'}\ e_1\ e_2) \sqsubseteq (\text{mon}^\ell_{\ell'}\ e_1'\ e_2')} \qquad \frac{\textit{free}(e^\bullet) \subseteq \{x\}}{(\lambda\ (x)\ e^\bullet) \sqsubseteq_F \bullet} \qquad \overline{n \sqsubseteq_F \bullet}$$

Fig. 22. Approximation between expressions

$$\overline{n \sqsubseteq_F n} \qquad \overline{op \sqsubseteq_F op} \qquad \frac{e \sqsubseteq_F e' \quad \rho \sqsubseteq_F \rho' \quad \phi \sqsubseteq_F \phi'}{\text{Clo}(x, e, \rho, \phi) \sqsubseteq_F \text{Clo}(x, e', \rho', \phi')}$$

$$\frac{\textit{free}(e^\bullet) \subseteq \{x\} \quad \text{restricted}_F(\rho)}{\text{Clo}(x, e^\bullet, \rho, \phi) \sqsubseteq_F \bullet} \qquad \frac{v \sqsubseteq_F v' \quad s \sqsubseteq s'}{(v, s) \sqsubseteq_F (v', s')} \qquad \overline{s \sqsubseteq \varnothing} \qquad \overline{s \sqsubseteq s}$$

Fig. 23. Approximation between runtime values

- Case $\mathcal{E}[(x, \rho)] \sqsubseteq_F \mathcal{E}'[(x, \rho')]$: Next states continue to approximate by rule *[Var]*, and existing approximation between environments, store-caches, and stores.
- Case $\mathcal{E}[(n, \rho)] \sqsubseteq_F \mathcal{E}'[(\bullet, \rho')]$: Next states step by rule [Lit]. The new states approximate because $n \sqsubseteq_F \bullet$.
- Case $\mathcal{E}[(\text{set!}\ (x, \rho)\ w)] \sqsubseteq_F \mathcal{E}'[(\text{set!}\ (x, \rho')\ w')]$: Next states continue to approximate by rule *[Set]*.
- Case $\mathcal{E}[(\text{if}\ w\ c_1\ c_2)] \sqsubseteq_F \mathcal{E}'[(\text{if}\ w'\ c_1'\ c_2')]$: By soundness of relation feasible? and that $w \sqsubseteq_F w'$, RHS must at least reduce through the rule that applies to LHS (either *[CondTrue]* or *[CondFalse]*), which preserves approximation.
- Case $\mathcal{E}[(w_1\ w_2)^\ell] \sqsubseteq_F \mathcal{E}'[(w_1'\ w_2')^\ell]$:
  - If $w_1'$ is concrete, both states must reduce through the same reduction rule and the next states preserve the approximation relation.
  - If $w_1'$ is $(\bullet, s)$, $w_1$ must contain purely instantiated code by the definition of $w_1 \sqsubseteq w_1'$. By assumption, $w_1$'s environment $\rho$ only has access to "leaked" values approximated by those at address $\alpha_\bullet$. The execution of $(w_1\ w_2)$ now has access to $w_2$ in addition, which is soundly approximated by rule *[AppOpq]* extending $\alpha_\bullet$ to containt the approximating value $w_2'$. (If $\alpha$ is the new address pointing to the value at $w_2$, the new abstraction map is $F[\alpha \mapsto \alpha_\bullet]$). The opaque application with store extended at $\alpha_\bullet$ continues to approximate the arbitrary state that $(w_1\ w_2)$ steps to.
- Case $(\mathcal{E}[c], m, \phi, \sigma) \sqsubseteq_F (\mathcal{E}'[(\bullet\ [w])], m', \phi', \sigma')$ because restricted$_F(c)$ and $\mathcal{E} \sqsubseteq_F \mathcal{E}'[(\bullet\ [])]$:

Soft Contract Verification for Higher-Order Stateful Programs 51:29

$$\frac{}{\{\} \sqsubseteq_F \{\}} \quad \frac{\rho \sqsubseteq_F \rho'}{\rho[x \mapsto \alpha] \sqsubseteq_F \rho'[x \mapsto F(\alpha)]} \quad \frac{m \sqsubseteq_F m' \quad w \sqsubseteq_F w'}{m[x \mapsto w] \sqsubseteq_F m'[x \mapsto w']}$$

$$\frac{c_1 \sqsubseteq_F c_1' \quad c_2 \sqsubseteq_F c_2'}{\mathcal{E}[(\mathsf{if}\,[\,]\,c_1\,c_2)] \sqsubseteq_F \mathcal{E}'[(\mathsf{if}\,[\,]\,c_1'\,c_2')]} \quad \frac{c \sqsubseteq_F c' \quad \mathcal{E} \sqsubseteq_F \mathcal{E}' \quad \ell \neq \ell_\bullet}{\mathcal{E}[([\,]\,c)^\ell] \sqsubseteq_F \mathcal{E}'[([\,]\,c')^\ell]}$$

$$\frac{w \sqsubseteq_F w' \quad \mathcal{E} \sqsubseteq_F \mathcal{E}' \quad \ell \neq \ell_\bullet}{\mathcal{E}[(w\,[\,])^\ell] \sqsubseteq_F \mathcal{E}'[(w'\,[\,])^\ell]} \quad \frac{\mathcal{E} \sqsubseteq_F \mathcal{E}' \quad \rho \sqsubseteq_F \rho'}{\mathcal{E}[(\mathsf{set!}\,(x,\rho)\,[\,])] \sqsubseteq_F \mathcal{E}'[(\mathsf{set!}\,(x,\rho')\,[\,])]}$$

$$\frac{\mathcal{E} \sqsubseteq_F \mathcal{E}' \quad m \sqsubseteq_F m'}{\mathcal{E}[(\mathsf{rt}_x^{\vec{y}}\,s\,m\,\phi\,[\,])] \sqsubseteq_F \mathcal{E}'[(\mathsf{rt}_x^{\vec{y}}\,s\,m'\,\phi\,[\,])]} \quad \frac{\mathcal{E} \sqsubseteq_F \mathcal{E}' \quad c \sqsubseteq_F \mathcal{E}' \quad \ell_\bullet \notin \{\ell,\ell'\}}{\mathcal{E}[(\mathsf{mon}_{\ell'}^\ell\,[\,]\,c)] \sqsubseteq_F \mathcal{E}'[(\mathsf{mon}_{\ell'}^\ell\,[\,]\,c')]}$$

$$\frac{\mathcal{E} \sqsubseteq_F \mathcal{E}' \quad w \sqsubseteq_F w' \quad \ell_\bullet \notin \{\ell,\ell'\}}{\mathcal{E}[(\mathsf{mon}_{\ell'}^\ell\,w\,[\,])] \sqsubseteq_F \mathcal{E}'[(\mathsf{mon}_{\ell'}^\ell\,w'\,[\,])]} \quad \frac{\mathcal{E} \sqsubseteq_F \mathcal{E}'}{\mathcal{E} \sqsubseteq_F \mathcal{E}[(\bullet\,[\,])]}$$

$$\frac{\mathsf{restricted}_F(c) \quad \mathcal{E} \sqsubseteq_F \mathcal{E}'[(\bullet\,w')^\ell]}{\mathcal{E}[([\,]\,c)^{\ell_\bullet}] \sqsubseteq_F \mathcal{E}'[(\bullet\,w')^\ell]} \quad \frac{\mathsf{restricted}_F(w) \quad \mathcal{E} \sqsubseteq_F \mathcal{E}'[(\bullet\,w')^\ell]}{\mathcal{E}[(w\,[\,])^{\ell_\bullet}] \sqsubseteq_F \mathcal{E}'[(\bullet\,w')^\ell]}$$

$$\frac{\mathsf{restricted}_F(c_1) \quad \mathsf{restricted}_F(c_2) \quad \mathcal{E} \sqsubseteq_F \mathcal{E}'[(\bullet\,w')^\ell]}{\mathcal{E}[(\mathsf{if}\,[\,]\,c_1\,c_2)] \sqsubseteq_F \mathcal{E}'[(\bullet\,w')^\ell]}$$

$$\frac{\mathsf{restricted}_F(\rho) \quad \mathcal{E} \sqsubseteq_F \mathcal{E}'[(\bullet\,w')^\ell]}{\mathcal{E}[(\mathsf{set!}\,(x,\rho)\,[\,])] \sqsubseteq_F \mathcal{E}'[(\bullet\,w')^\ell]} \quad \frac{\sigma \sqsubseteq_F \sigma' \quad v \sqsubseteq_F v'}{\sigma \sqcup [\alpha \mapsto v] \sqsubseteq_F \sigma' \sqcup [F(\alpha) \mapsto v']}$$

$$\frac{c \sqsubseteq_F c' \quad m \sqsubseteq_F m' \quad \phi \sqsubseteq_F \phi' \quad \sigma \sqsubseteq_F \sigma}{(c,m,\phi,\sigma) \sqsubseteq_F (c',m',\phi',\sigma')}$$

$$\frac{\mathsf{restricted}_F(c) \quad \mathcal{E} \sqsubseteq_F \mathcal{E}'[(\bullet\,[\,])] \quad m \sqsubseteq_F m' \quad \phi \sqsubseteq_F \phi' \quad \sigma \sqsubseteq_F \sigma'}{(\mathcal{E}[c],m,\phi,\sigma) \sqsubseteq_F (\mathcal{E}'[(\bullet\,w)],m',\phi',\sigma')}$$

Fig. 24. Approximation between machine components

Either the same non-structural approximation continues to hold between RHS and LHS's next state, or if LHS transfers control to transparent code by applying a function, the function must be approximated by one value in $\sigma'(\alpha_\bullet)$, which *[AppOpq]* soundly approximates by non-deterministically applying one.

□

THEOREM A.2 (BLAME SOUNDNESS). *If $e_1 \sqsubseteq e_1'$ and $\mathsf{load}(e_1) \longmapsto (\mathcal{E}[\mathsf{blame}_{\ell'}^\ell], m, \phi, \sigma)$, where $\ell \neq \ell_\bullet$, then there exists $\mathcal{E}'$, $m'$, $\phi'$, and $\sigma'$, such that $\mathsf{load}(e_1') \longmapsto (\mathcal{E}'[\mathsf{blame}_{\ell'}^\ell], m', \phi', \sigma')$.*

PROOF. The proof proceeds by rule-induction on the derivation of $(\longmapsto)$ in the concrete error trace. The base case (reflexive) vacuously holds. The inductive case (transitive) holds by lemma 3.1, where for each single reduction step ($\longmapsto$) on the concrete state, the abstract state continues to approximate the concrete state in zero or more steps. □

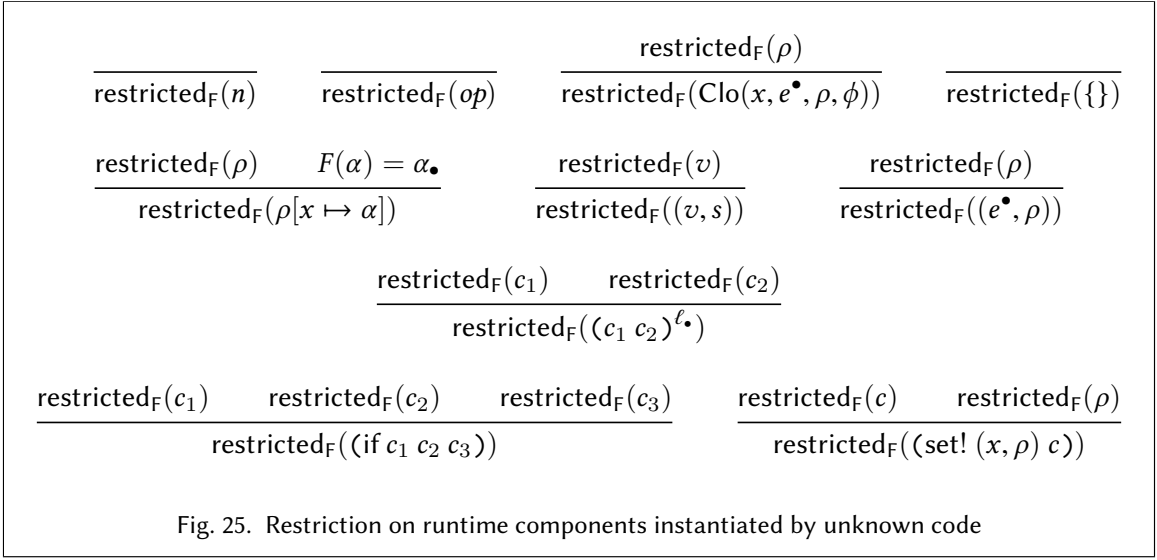

Fig. 25. Restriction on runtime components instantiated by unknown code